\documentclass[conference]{IEEEtran}
\usepackage{cite}
\usepackage{amsmath,amssymb,amsfonts}
\usepackage{algorithmic}
\usepackage{graphicx}
\usepackage{textcomp}
\usepackage{xcolor}
\def\BibTeX{{\rm B\kern-.05em{\sc i\kern-.025em b}\kern-.08em
    T\kern-.1667em\lower.7ex\hbox{E}\kern-.125emX}}
\begin{document}

\title{Developing Assurance Cases for Adversarial Robustness and Regulatory Compliance in LLMs}

\author{
\IEEEauthorblockN{Tomas Bueno Momcilovic}
\IEEEauthorblockA{
\textit{fortiss research institute}\\
Munich, Germany \\
momcilovic@fortiss.org
}
\hspace{5cm}
\and
\IEEEauthorblockN{Dian Balta}
\IEEEauthorblockA{
\textit{fortiss research institute}\\
Munich, Germany \\
}
\hspace{5cm}
\and
\IEEEauthorblockN{Beat Buesser}
\IEEEauthorblockA{\textit{IBM Research Europe}\\
Zurich, Switzerland \\
}
\hspace{5cm}
\and
\IEEEauthorblockN{Giulio Zizzo}
\IEEEauthorblockA{\textit{IBM Research Europe}\\
Dublin, Ireland \\
}
\and
\IEEEauthorblockN{Mark Purcell}
\IEEEauthorblockA{\textit{IBM Research Europe}\\
Dublin, Ireland \\
}
\and 
\IEEEauthorblockN{    }
\IEEEauthorblockA{\textit{     } \\
\textit{     }\\
 \\
   }
}

\maketitle

\begin{center}
    \textbf{Accepted manuscript.}
\end{center}

\begin{abstract}
This paper presents an approach to developing assurance cases for adversarial robustness and regulatory compliance in large language models (LLMs). Focusing on both natural and code language tasks, we explore the vulnerabilities these models face, including adversarial attacks based on jailbreaking, heuristics, and randomization. We propose a layered framework incorporating guardrails at various stages of LLM deployment, aimed at mitigating these attacks and ensuring compliance with the EU AI Act. Our approach includes a meta-layer for dynamic risk management and reasoning, crucial for addressing the evolving nature of LLM vulnerabilities. We illustrate our method with two exemplary assurance cases, highlighting how different contexts demand tailored strategies to ensure robust and compliant AI systems.
\end{abstract}


\section{Introduction}

As the deployment of large language models (LLMs) becomes increasingly widespread, their vulnerability to adversarial attacks has emerged as a significant concern. These attacks involve crafting inputs that bypass the models' safety mechanisms, leading to the generation of harmful outputs. Traditional adversarial attacks in machine learning often rely on making subtle, nearly undetectable modifications to input data. However, in the context of LLMs, the nature of these attacks evolves to include more sophisticated strategies such as gradient-based optimizations, persuasive tactics that circumvent established guardrails, and model inversions that can produce vulnerable code. This complexity necessitates a robust framework to ensure that LLMs remain secure and reliable in their intended applications.

In parallel, regulatory frameworks like the EU AI Act introduce new compliance requirements that LLM developers and deployers must meet, particularly concerning adversarial robustness. The Act, which categorizes LLMs as general-purpose AI, imposes a variety of obligations on developers of systems deploying LLMs in particular contexts, including the need to safeguard against adversarial attacks and report serious incidents. The dynamic nature of LLM vulnerabilities and repeated interactions with the application make the prediction and prevention of such incidents an extremely difficult and continuous effort.

For these reasons, there is a pressing need for a structured approach to represent and reason about adversarial attacks and guardrails. Assurance practices (cf. \cite{hawkins2021}) provide an important foundation towards satisfying that need. We propose one approach to developing assurance cases, with the goal to primarily support system developers, security engineers and auditors, but also LLM developers, to reason and report on robustness and compliance of their applications. These cases not only address the quantifiable aspects of adversarial robustness or checklist-based compliance, but also ensure that qualitative and intermediate strategies - or their technical or legal defeaters - are appropriately represented. We instantiate the assurance cases in examples on natural language and computer language (i.e., coding) tasks.

\section{Background}

\subsection{Adversarial Attacks}

Traditional adversarial attacks in machine learning often involve making small, nearly imperceptible changes to the input data, to mislead a model into making incorrect predictions \cite{carlini2017adversarial}. By contrast, adversarial attacks on large language models (LLMs) typically involve crafting prompts to bypass the model's guardrails and generate outputs that are harmful \cite{Zou2023_Universal,Zeng2024_How}.

Attacks can be executed using gradient-based input modifications, persuasive tactics to circumvent guardrails, or model inversion to produce harmful code \cite{hajipour2023systematically}. They often utilize techniques like jailbreaking, where manually crafted inputs exploit model vulnerabilities, heuristic optimization through semi-automated inputs leveraging learned properties, or randomization, which involves automatically generated inputs targeting the model's coverage gaps.

For the purpose of this paper, an adversarial attack is operationalized as any input that may lead to output that is unintended\footnote{Unintended output can include, for example, hazardous, harmful, toxic, hallucinated or deceptive content.} by either LLM developer, system developer or both. This definition includes both malicious attacks and accidental model failures. Because motive can often be ambiguous, an adversary can be an attacker with malicious intent, but also a benign user whose input is adversarial by accident or out of curiosity (i.e., red-teaming or white-hat hacking), and any other entity (e.g., bot).

\subsection{Guardrails and Layering}

Guardrails against adversarial attacks are varied and rapidly evolving. They are primarily focused on detecting adversarial input or unintended model output. For example, one set of guardrails centers on filtering input based on perplexity, which is an estimate of how "surprised" is an LLM with a particular prompt \cite{alon2023detecting}. Another example are guardrails which identify jailbreak attempts by detecting erratic output \cite{robey2023smoothllm}. Developers with white-box access to the LLM can also train the model to internalize guardrails, such that the LLM is fine-tuned to autonomously recognize and reject harmful prompts \cite{helbling2023llm}. However, the interpretability of models, especially with regard to why certain attacks succeed, is a known challenge.

\begin{figure*}
    \centering
    \includegraphics[width=\linewidth]{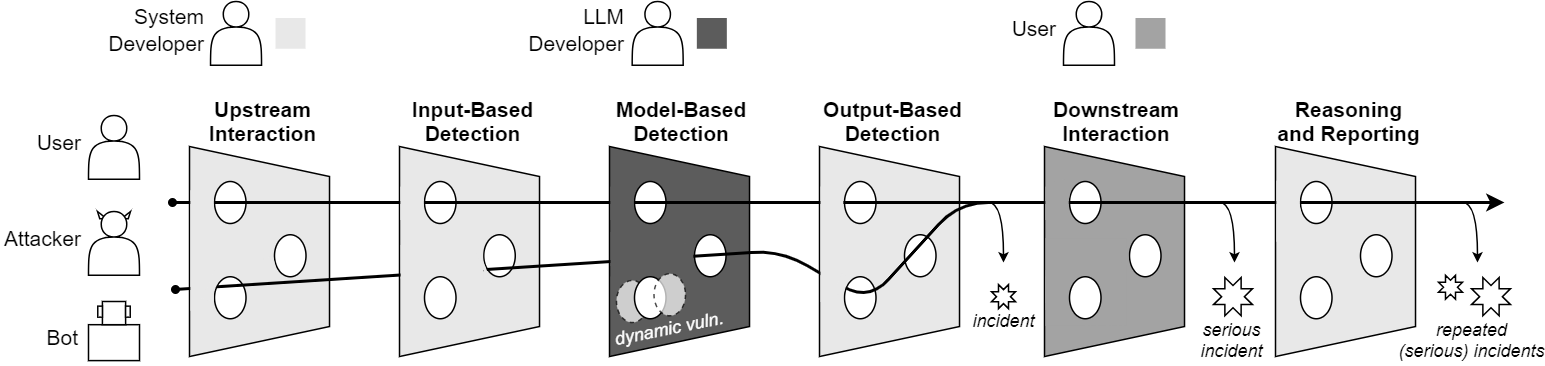}
    \caption{Swiss cheese diagram as a conceptual and generic representation of layers, vulnerabilities and roles concerned with deployed LLM-based application.}
    \label{fig:swiss_cheese}
\end{figure*}

We focus on the black-box, conversational and pretrained LLMs whose deployment can range from simple user-facing chatbots, to complex pipelines with up- and downstream components (i.e., compound AI systems \cite{compound-ai-blog}). The functionally simplest yet useful application would require only two layers: an interface for gathering input from, and displaying output to, a user; and a model for processing input and generating output. However, the success rate of simple variations of known attack patterns \cite{geiping2024} shows that relying purely on LLMs is not enough to claim robustness. Therefore, following the practices of secure software engineering, an adversarially robust application may require further layers.

We observe five layers which can incorporate guardrails (cf. \ref{fig:swiss_cheese}). The first layer, the interface for upstream interaction, includes mechanisms that constrain or deter potential adversaries before they submit an input prompt. Examples of such mechanisms include access controls, challenge-response tests (e.g., CAPTCHAs) or character limitations, but also instructions, disclaimers, and warnings that alert unintentional adversaries. 

The second layer for detection of adversarial input, consists of guardrails that constrain potential attacks after the prompt is submitted to the application, but before it reaches the LLM. This can include detection-based input filters, system instructions provided to the LLM, prompt preprocessing, or even advanced classifiers using specialized models (i.e., LLM-as-a-Judge \cite{zheng2023judgingllm}).

The third layer is the LLM itself, or rather the mechanisms inside the model's black box that detect anomalies and respond accordingly, or reduce the incidence of harmful output. Techniques such as adversarial training, grounding, unlearning, and reinforcement learning with human feedback (RLHF) can be used to set up such guardrails \cite{wang2023self}. 

The fourth layer for detection of unintended output, focuses on mechanisms that constrain primarily harmful content from reaching the user. These can include keyword-based filters, postprocessing, human-in-the-loop systems, or again, the LLM-as-Judge approach. 

The fifth and final layer for interaction with downstream components or users, is concerned with reducing the consumption of harmful output even after it is provided. This could involve providing additional information such as the model's confidence level or instructions that support critical thinking (i.e., AI literacy), or components and tools that allow users and other applications to verify structured output, flag errors, or simulate execution in sandbox environments \cite{ruan2024identifyingrisks}.

\section{Challenges in Establishing Compliant Robustness}

\subsection{EU AI Act and Incidents}

The EU AI Act \cite{euaia2024corrigendum} is one of the first regulations to deal with AI systems, referring to LLMs as general-purpose AI. The Act regulates in which domains and products (i.e., contexts) can AI be introduced and how, but does not regulate the models themselves. Most of the obligations are placed on providers of AI systems, such as system developers who deploy LLMs in such regulated but non-prohibited contexts (cf. \cite{novelli2024generative}). Unless they are the same entity as system developers, the LLM developers of models made available in the EU have obligations that are largely limited to factsheets.

At least 20 identifiable duties \cite{bueno2024assuring} concern the (cyber)security, safety and robustness of AI systems. One such duty places an explicit demand for protection from adversarial attacks (Article 15 Paragraph 5, \cite{euaia2024corrigendum}). However, other duties link implicitly to the attacks through their potential consequences. One such duty identifies serious incidents\footnote{Serious incidents are defined as "[any] incident or malfunctioning leading to death or serious damage to health, serious and irreversible disruption of the management and operation of critical infrastructure, infringements of obligations under Union law intended to protect fundamental rights or serious damage to property or the environment." (Article 3 Point 49 \& Recital 155, EUAIA)} resulting from AI systems, and demands from system developers to report their occurence to supervisory authorities (Article 73, EUAIA). Another duty refers to more ambiguous systemic risks\footnote{Systemic risk is a "risk that is specific to the high-impact capabilities of general-purpose AI models, having a significant impact on the Union market due to their reach, or due to actual or reasonably foreseeable negative effects on public health, safety, public security, fundamental rights, or the society as a whole, that can be propagated at scale across the value chain." (Article 3 Point 65, EUAIA)}, whereby the developers of general-purpose AI embodying such risks are obliged to fulfill additional requirements.

Here, we operationalize any unintended output that is provided to the user as a \textit{de facto} incident from the point-of-view of system and LLM developers, regardless of the downstream impact or the user's motives. As visualized in Figure \ref{fig:swiss_cheese}, an incident becomes serious if the output is used downstream in a way that leads to the consequences defined in the Act, which likely includes those of a systemic nature. The opposite can also be inferred, such that a given incident would generally not become serious if the output is ignored or its effect is successfully prevented, even if its potential impact is high\footnote{However, even if a potentially serious incident did not lead to serious consequences thanks to a downstream guardrail, it is currently not clear whether the repeated occurence of such an incident could be interpreted as proof of deficient risk management or systemic risk.}.

\subsection{Dynamic Vulnerabilities and Repeated Use}

Given this operational definition, we can infer that incidents can be expected. Two premises support this claim: the dynamicity of the vulnerabilities, and the repeated and extensible nature of interactions with LLMs.

Vulnerabilities of LLMs are virtually dynamic, even if they are fundamentally static for pretrained LLMs without further changes, because the attack surface cannot be approximated for at least three reasons. First, logging and white-box analyses are resource-intensive and provide limited interpretability of model behavior due to the size and density of language models. Second, system developers who opt for pretrained LLMs from external sources, in practice have only black-box or limited access to the model architecture. Third, the coverage of guardrails is not easily known beforehand, considering the possible combinations and sequences of allowable input and output, model probabilities and parameters, and downstream contexts and components.

Additionally, LLMs have proven to be effective for repeated use and highly extensible. Most LLM architectures can leverage some form of memory, and perform well in tasks beyond language translation, inviting sequential and cross-purpose use. An LLM instance in the initial state or one domain may be significantly different from the instances in subsequent states or other domains. If users or downstream components depend on previous memory or multiple interactions to generate useful output \cite{shi2022}, or if LLM-based applications are designed to be extensible, the surface area for adversarial attacks increases, allowing for greater exploitation (e.g., through multi-turn attacks \cite{russinovich2024great}).

It is thus important to understand how to handle incidents continuously, both from the robustness and the compliance perspective. Practical adversarial robustness requires continuous monitoring for novel attacks, understanding how models respond to different patterns, and operationalizing this knowledge in the guardrails. This continuity is also recognized in the Act's references to quality and risk management, and the necessity of investigation into the causes of serious incidents.

We represent this continuity as a meta-layer that oversees the management of incidents and guardrails over time. The additional sixth layer, reasoning and reporting, is an umbrella for handling dynamic risk. The reasoning aspect centers on the techniques that allow developers to evaluate guardrail performance over time, including counterfactual evaluation, early warning systems, or anomaly detection, and define rules or policies based on context and attack patterns. The reporting aspect supports the developers in fulfilling their duties to other stakeholders, but also gathering information. The former includes factsheets with metrics, benchmarks and verifiable tests, distributed warnings or information from investigating serious incidents. The latter includes the collection of feedback, reports of harmful content and bug reports from the users. This plays a crucial role in ensuring stakeholders are accountable for incidents. Vulnerability in reasoning and reporting thus means that certain incidents go unnoticed or unaddressed.

We find assurance cases to be the fundamental components of this meta-layer, and explore their role in the next section.

\section{Assurance Case for Ensuring Robustness}

Assurance cases have been shown to be suitable to creating arguments for assuring different properties of machine learning components \cite{ishikawa2018argument,gallina2023}. Inspired by such work, we create assurance cases that can be instantiated to feasibly cover both compliance and robustness. The structured language of assurance arguments provides the basis for machine-understandable reasoning, while the graphical notation provides the human-readable report. Combined with a simple engine (e.g., when encoded in an ontology and stored in a graph database), the argument can be used to evaluate and express robustness and compliance in explicit terms.

\begin{figure*}[hbt!]
    \centering
    \includegraphics[width=0.85\linewidth]{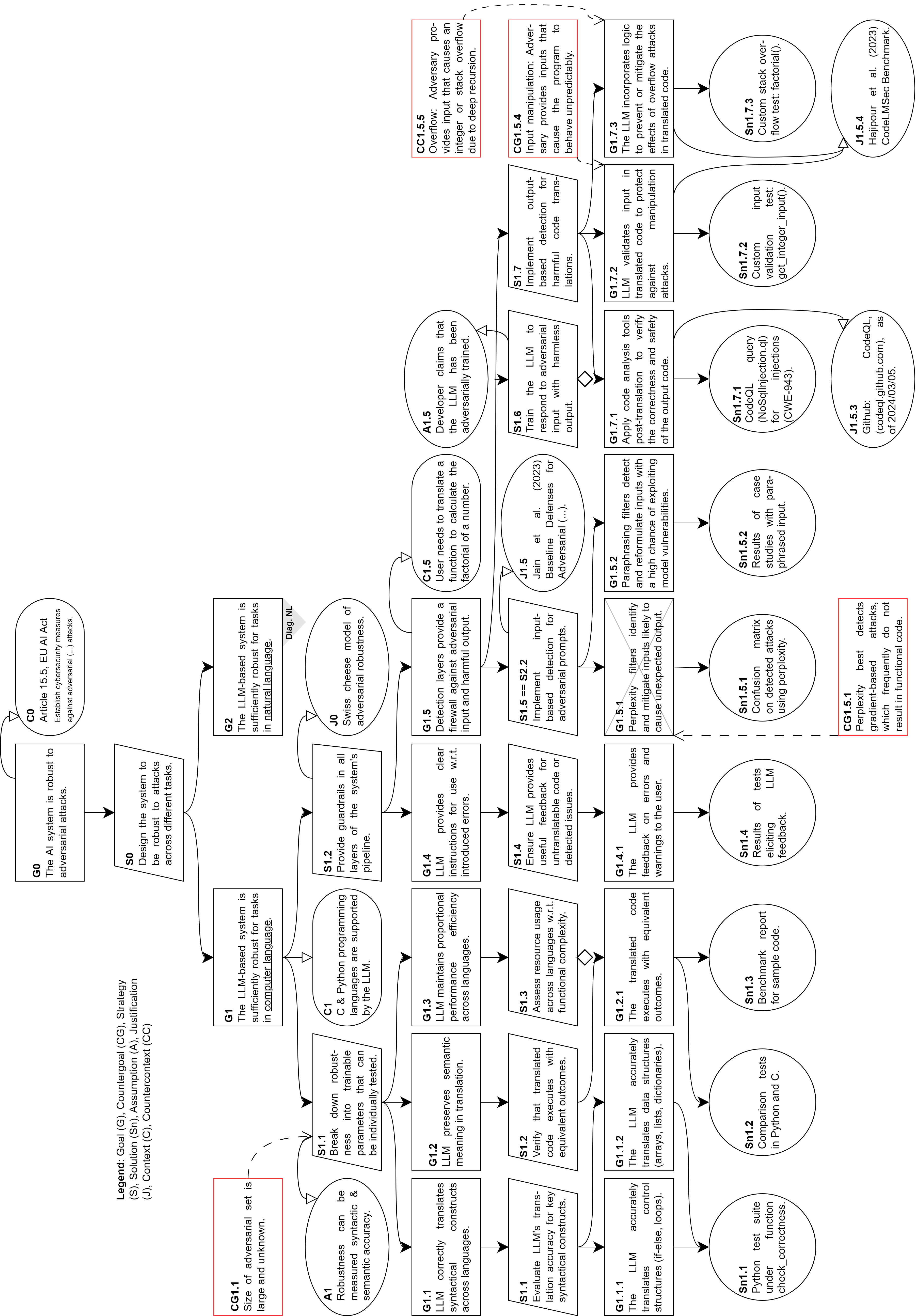}
    \caption{Exemplary assurance case for LLMs used in code language tasks.}
    \label{fig:assurance_case_code}
\end{figure*}

\begin{figure}[hbt!]
    \centering
    \includegraphics[width=\linewidth]{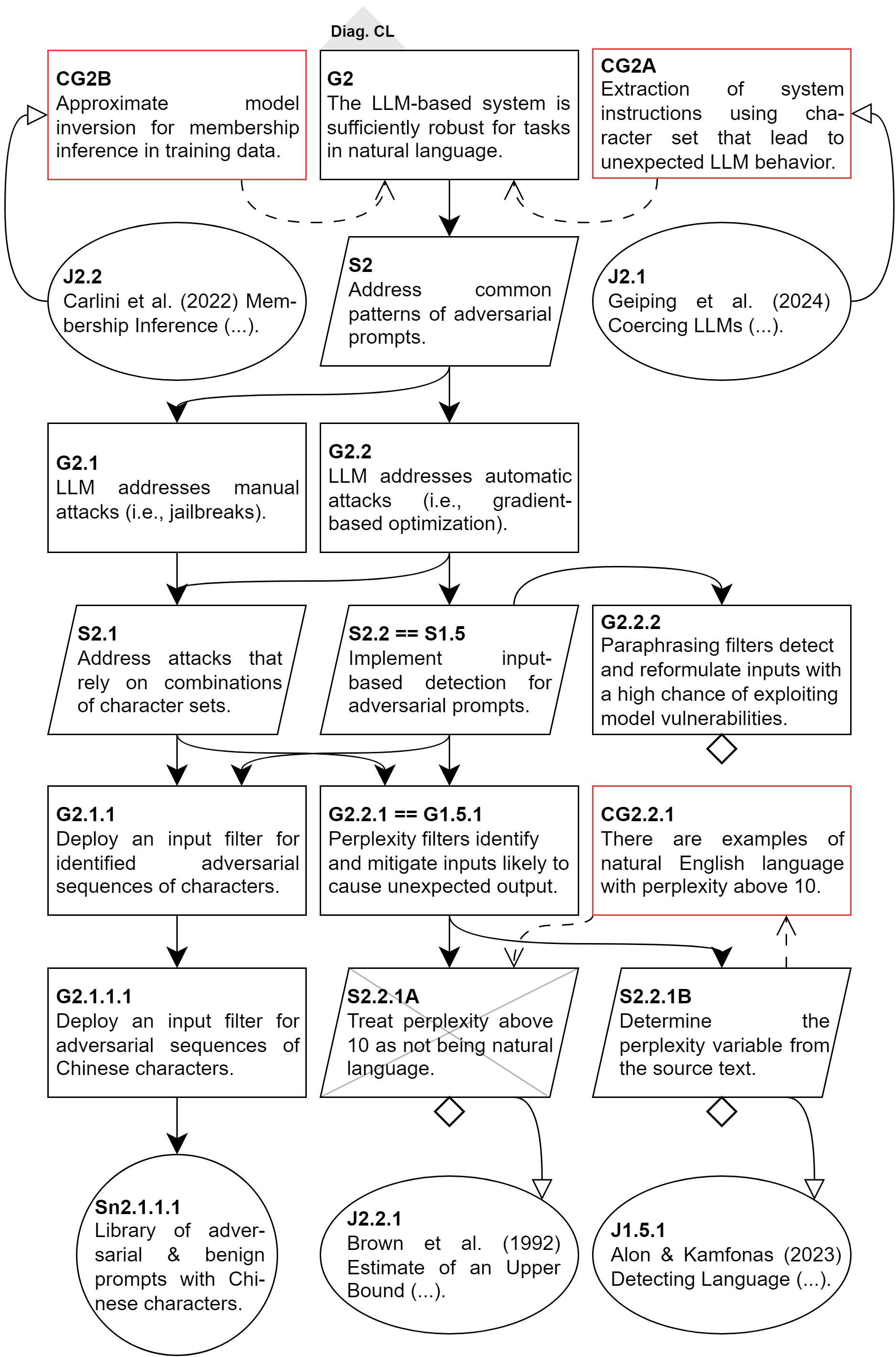}
    \caption{Exemplary assurance case for LLMs used in natural language tasks.}
    \label{fig:assurance_case_lang}
\end{figure}

The presented assurance cases in Figures \ref{fig:assurance_case_code} and \ref{fig:assurance_case_lang} have been created following the Goal Structuring Notation and its corresponding best practices \cite{acwg2021}. Each case includes a subset of relevant vulnerabilities and adversarial attacks, contingent on the context and tasks for which the LLM is used. Each claim is a result of a dialectical discussion and a review of common attack patterns (cf. \cite{Zou2023_Universal,geiping2024}).

For tasks related to natural languages, such as text translation, generation, or autocompletion, the LLM is vulnerable to a broad spectrum of both manually and automatically generated prompts that may lead to harmful output. This harmful output can affect users, downstream recipients, or any potential audience. For instance, toxic or derogatory remarks can cause harm to readers, while providing helpful instructions related to dangerous activities presents an information hazard. Protecting against these issues may involve identifying input and output patterns, such as correlated keywords that are commonly associated with toxic, harmful, or out-of-distribution text. However, this task is challenging due to the inherent complexity of language.

In contrast, for tasks related to computer languages, such as code translation, generation, review, or autocompletion, the LLM may be vulnerable to similar types of attacks. However, aside from any generated text, the harmfulness of the code output largely depends on how the code is applied downstream. There is a clear distinction between harmful functional code, vulnerable but functional code, and non-functional code—the latter two are only harmful if the user applies them in a downstream context. Due to the more deterministic nature of computer languages, it might be easier for the input- and output-detection layers to prevent harmful or vulnerable code from reaching the user. In cases where the user's intent behind a prompt is unclear, downstream protection could include making the user aware of vulnerabilities in the code, such as injection flaws, and suggesting potential remedies, such as implementing tests and input sanitization.

Therefore, exploitable vulnerabilities and serious incidents vary depending on the task context of the LLM. We address what compliant adversarial robustness means in each context respectively.

\subsection{Case 1: Computer Languages}

The assurance case for computer languages focuses on ensuring that LLMs used in code-related tasks can be relied on. The central objective is to make the AI system robust to adversarial attacks (G0). The assurance is structured around the idea that the LLM-based system is designed to be resilient across all programming languages it supports (G1), acknowledging that robustness is particularly important in these contexts due to their deterministic nature (C1.2).

The robustness of the LLM is linked to its ability to preserve the semantic integrity of code during translation (C1.1). The approach to ensuring this robustness involves breaking down potential vulnerabilities into specific programming contexts and implementing targeted guardrails across all tasks (S1). To maintain functional correctness and syntactical structure in the translated code, specific tests are performed (Sn1.1), ensuring that there is no semantic drift during translation (C1.4). This process includes using custom libraries to detect common vulnerabilities, such as SQL injection, in the translated code (Sn1.2).

However, the assurance case also identifies challenges that must be addressed, such as the risk of adversaries exploiting context switching or introducing complex edge cases that the guardrails might not fully account for (CC1.5.5). This highlights the need for more refined detection and mitigation strategies within the system.

The reasoning here relies on compositional and defeasible nature of guardrails for software. For a Python-to-C subset of all code translation tasks, for example, software tests would provide measurable coverage at input, output and downstream layers. As the LLM is increasingly prompted with various source code, coverage can become the main rule for proactively generating defeaters to previously valid goals (CG1.5.1), and deprecating guardrails that are no longer needed (G1.5.1).

\subsection{Case 2: Natural Languages}

In this assurance case, the focus is on ensuring that LLMs are robust when handling tasks in natural language processing (NLP). The system’s robustness is aimed at handling a variety of tasks effectively (G2). However, it is recognized that extracting system instructions from character sets can sometimes lead to unexpected LLM behavior, which is a potential vulnerability (CG2A).

To address these vulnerabilities, the assurance case outlines that the system must be capable of mitigating both manual attacks, such as jailbreaking (G2.1), and automatic attacks, such as those involving gradient-based optimization (G2.2). The strategy involves focusing on common patterns of adversarial prompts (S2). This includes identifying and filtering specific character sets and sequences that could be exploited in attacks (S2.1) and employing perplexity filters to detect and mitigate inputs likely to cause unexpected outputs (G2.2.1). Additionally, paraphrasing filters are used to reformulate inputs that have a high chance of exploiting model vulnerabilities (G2.2.2).

Despite the guardrails, challenges such as approximate model inversion are likely to remain (CG2B). This underscores the need for ongoing research and development to enhance the robustness of LLMs in NLP tasks. For a given system developer, the evasive attack surface of their system needs to be handled with defeasible and updateable claims and evidence. A developer can implement a guardrail (S2.2.1A) based on a naive best-estimate prior (J2.2.1), but maintaining the argument over time means that the errors and vulnerabilities of patches are properly represented (CG2.2.1) and handled (S2.2.1B) using new information (J1.5.1). In cases where such information cannot be integrated without involving the LLM developer to retrain the model, but attack patterns can be observed (G2.1.1), the system developer add constraints (G2.1.1.1) that at least limit the surface area, until an appropriate defeater and updated strategy can be formulated.

\newpage

\section{Concluding Remarks and Further Work}

Our paper provides an overview of what robustness to adversarial attacks means, how can it be argued to be compliant with the EU AI Act, and how assurance argumentation provides the crucial component for proving compliant adversarial robustness. With this, we provide the following remarks.

First, we posit that both the regulation and the nature of LLMs make adversarial robustness dependent on the composition of different guardrails. We structure these guardrails into layers according to a standard LLM-based application pipeline. However, we also argue that there is a need for a meta-layer to manage these guardrails with some dynamicity, without which there is no assurance.

Second, we posit that successful attacks will occur. Thus, the strength of the assurance case lies in how well the system deals with such challenges. This is the core duty of the Act, whereby serious incidents must also lead to reporting and investigation. This also means that the assurance case needs to be defined dialectically, incorporating new attacks and re-evaluating guardrails.

Third, our work is currently limited to pretrained LLMs that are not trained further on new data. Continuous training would introduce a substantial change, such that any reasoning about guardrail coverage and effectiveness would require new evidence. Finally, our understanding of incidents is presented only for the purposes of guiding the development of the assurance case. Whether unintentional output can also include false positives, which prevent the user or component from receiving critical answers to otherwise legitimate prompts, is open to interpetation (e.g., being denied CPR instructions when performing first-aid).

Our further work in this area centers on providing the fundament for such a meta-layer. We focus on establishing dynamicity of reasoning with and through assurance cases, by using ontologies and graph databases for querying. As we plan to evaluate such a system, we hope to support the system developers, LLM developers and auditors in ensuring their systems are adversarially robust, and compliant at that.

\section*{Acknowledgment}
This work was partially supported by financial and other means by the following research projects: DUCA (EU grant agreement 101086308), FLA (supported by the Bavarian Ministry of Economic Affairs, Regional Development and Energy), the DiProLeA (German Federal Ministry of Education and Research, grant 02J19B120 ff), as well as our industrial partners in the FinComp project. We thank the reviewers for their valuable comments.

\bibliographystyle{IEEEtran}
\bibliography{refs}

\end{document}